\documentclass[journal]{IEEEtran}
\usepackage{amsmath}
\usepackage{algorithm,algpseudocode,lipsum}

\usepackage{epstopdf} 
\usepackage{tabularx,threeparttable}
\usepackage{array}
\usepackage{amssymb, amsmath, bm}
\usepackage{mathtools}
\usepackage{mathrsfs}
\usepackage{booktabs}
\usepackage{multirow}
\usepackage{url}
\usepackage{color}
\usepackage{dsfont}
\usepackage{subcaption}
\usepackage{graphicx} 
\usepackage{caption}
\usepackage{subcaption}

\newcommand{\revise}[1]{#1}

\def\ie{\emph{i.e.}}
\def\eg{\emph{e.g.}}
\def\etal{{\em et al.}}

\newcolumntype{?}[1]{!{\vrule width #1}}

\usepackage{cite}

\begin{document}
	\title{Deep Sinogram Completion with Image Prior for Metal Artifact Reduction in CT Images}

	\author{Lequan Yu,
		Zhicheng Zhang, 
		Xiaomeng Li,
		and Lei Xing
		\thanks{Manuscript received on May 05 2020, revised on XX XX 2020.}
		\thanks{L. Yu, Z. Zhang, X. Li, and L. Xing are with the Department of Radiation Oncology, Stanford University, Palo Alto, CA 94306, USA (e-mail: \{lequany, zzc623, xmengli, lei\}@stanford.edu).}
		\thanks{This work was partially supported by NIH (1 R01CA227713), Varian Medical Systems, and a Faculty Research Award from Google Inc. }
		\thanks{Copyright (c) 2020 IEEE. Personal use of this material is permitted.
			Permission from IEEE must be obtained for all other uses, including
			reprinting/republishing this material for advertising or promotional purposes,
			collecting new collected works for resale or redistribution to servers or lists,
			or reuse of any copyrighted component of this work in other works. }
	}

	%\markboth{IEEE TRANSACTIONS ON MEDICAL IMAGING}%
	\markboth{IEEE TRANSACTIONS ON MEDICAL IMAGING}%
	{Shell \MakeLowercase{\textit{et al.}}: Bare Demo of IEEEtran.cls for IEEE Journals}

	\maketitle
	
	\begin{abstract}
	Computed tomography (CT) has been widely used for medical  diagnosis, assessment, and therapy planning and guidance.
	In reality, CT images may be affected adversely in the presence of metallic objects, which could lead to severe metal artifacts and influence clinical diagnosis or dose calculation in radiation therapy.
	In this paper, we propose a generalizable framework for metal artifact reduction (MAR) by simultaneously leveraging the advantages of image domain and sinogram domain-based MAR techniques.
	We formulate our framework as a sinogram completion problem and train a neural network (SinoNet) to restore the metal-affected projections.
	To improve the continuity of the completed projections at the boundary of metal trace and thus alleviate new artifacts in the reconstructed CT images, we train another neural network (PriorNet) to generate a good prior image to guide sinogram learning, and further design a novel residual sinogram learning strategy to effectively utilize the prior image information for better sinogram completion.
	The two networks are jointly trained in an end-to-end fashion with a differentiable forward projection (FP) operation so that the prior image generation and deep sinogram completion procedures can benefit from each other.
	Finally, the artifact-reduced CT images are reconstructed using the filtered backward projection (FBP) from the completed sinogram.
	\revise{Extensive experiments on simulated and real artifacts data} demonstrate that our method produces superior artifact-reduced results while preserving the anatomical structures and outperforms other MAR methods.
\end{abstract}


	\IEEEpeerreviewmaketitle

\section{Introduction}

\IEEEPARstart{C}{omputed} tomography (CT) systems have become an important tool for medical diagnosis, assessment, and therapy planning and guidance.
However, the metallic implants within the patients, \eg, dental fillings and hip prostheses, would lead to missing data in X-ray projections and cause strong star-shape or streak artifacts to the reconstructed CT images~\cite{de1998metal}.
Those metal artifacts not only present undesirable visual effects in CT images with influencing diagnosis but also make dose calculation problematic in radiation therapy~\cite{kalender1987reduction,meng2010sinogram}.
With the increasing use of metallic implants,  how to reduce metal artifacts has become an important problem in CT imaging~\cite{park2018ct}.

\begin{figure}[t]
	\centering
	\includegraphics[width=0.5\textwidth]{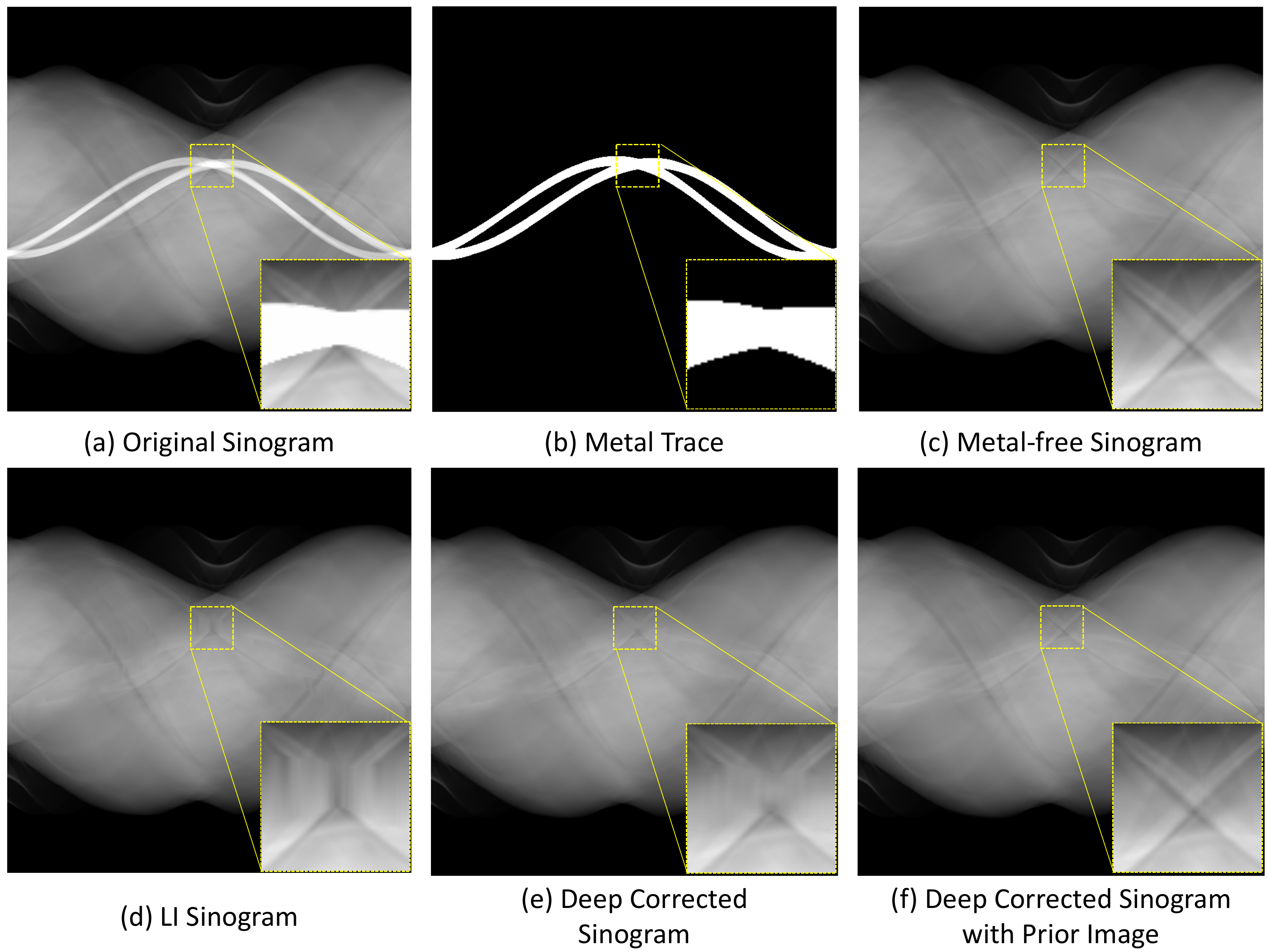}
	\caption{The qualitative  comparison of sinogram completion. An ROI is enlarged and shown with a narrower window to better visualize the difference.
	The linear interpolation (LI)~\cite{kalender1987reduction} produces a poor estimation of the missing projections (d), while the deep network can generate a relatively good corrected sinogram (e). With the guidance of prior image, our method predicts more accurate projections (f), which are very close to the metal-free one.} 
	\label{fig:sinocompletion}
\end{figure}

Numerous metal artifact reduction (MAR) methods have been proposed in the past decades, while there is no standard solution in clinical practice~\cite{meng2010sinogram,huang2015evaluation,gjesteby2016metal}.
Since the metal artifacts are structured and non-local in the reconstructed CT images, the previous metal artifact reduction approaches mainly addressed this problem in the X-ray projections (sinogram).
The metal-affected regions in the sinogram domain were corrected by modeling the underlying physical effects of imaging~\cite{hsieh2000iterative,kachelriess2001generalized,meyer2010empirical,park2015metal}.
For example, Park~\etal~\cite{park2015metal}  proposed a method to correct beam hardening artifacts caused by the presence of metal in polychromatic X-ray CT. 
However, with the presence of high-atom number metals, the metal trace regions in sinogram are often severely corrupted and the above methods are limited in achieving satisfactory results~\cite{zhang2018convolutional}. 
Therefore, the other MAR methods regarded the metal-affected regions as the missing areas and filled them with estimated values~\cite{kalender1987reduction,mehranian2013x}.
The early Linear interpolation (LI) approach~\cite{kalender1987reduction} filled the missing regions by the linear interpolation of its neighboring unaffected data for each projection view.
As interpolation cannot completely recover the metal trace information, the inconsistency between interpolated values and those unaffected values often results in strong new artifacts in the reconstructed images.
To improve the sinogram interpolation quality, recent methods involved the forward projection of a prior image to complete the sinogram~\cite{muller2009spurious,prell2009novel, meyer2010normalized, wang2013metal,zhang2013hybrid}.
These methods first estimated prior images with various tissue information from the uncorrected image and then performed forward projection on the prior image to conduct sinogram completion.
For example, Meyer \etal~\cite{meyer2010normalized} improved the LI approach by generating a prior image with tissue processing and normalizing the projection with a forward projection of the prior image before interpolation.
As the inaccurate prior images would lead to unfaithful structures in the reconstructed images, a key factor for prior-image-based approaches is to generate a good prior image to provide a more accurate surrogate for the missing data in the sinogram. 
Also, some researchers focused on designing new iterative reconstruction algorithms to reconstruct artifact-free images from the unaffected or corrected projections~\cite{wang1996iterative,wang1999iterative,lemmens2008suppression,zhang2011metal}.
For example, Zhang~\etal~\cite{zhang2011metal} proposed an iterative metal artifact reduction algorithm based on constrained optimization.
However, these iterative reconstruction methods often suffer from heavy computation and require proper hand-crafted regularizations. 

With the development of deep learning in medical image reconstruction and analysis~\cite{wang2018image,zhang2018sparse,litjens2017survey,ronneberger2015u},  recent progress of MAR has featured neural networks~\cite{park2018ct,gjesteby2017deep,zhang2018convolutional}.
Park~\etal~\cite{park2018ct} employed a U-Net~\cite{ronneberger2015u} in the sinogram domain to deal with beam-hardening related artifacts in polychromatic CT.
Gjesteby~\etal~\cite{gjesteby2017deep} utilized deep learning to refine the result of NMAR~\cite{meyer2010normalized} for achieving additional correction in critical image regions.
Zhang~\etal~\cite{zhang2018convolutional} proposed to generate a reduced-artifact prior image with CNN to help correct the metal-corrupted regions in the sinogram.
Although these methods show reasonable results on MAR, they are limited in handling with the remaining new artifacts in reconstructed CT images.

To improve the quality of the reconstructed CT images, inspired by the success of deep learning in solving ill-posed inverse problems in natural image processing~\cite{ledig2017photo,ulyanov2018deep,lehtinen2018noise2noise}, very recent works formulated MAR as an image restoration problem and reduced the metal artifacts with image-to-image translation networks~\cite{park2017machine,gjesteby2017reducing,gjesteby2017deep,gjesteby2018deep,liang2019metal,liao2019adn,lin2019dudonet,lyu2020dudonet}.
Gjesteby~\cite{gjesteby2018deep} employed a deep neural network to reduce the new artifacts after the  NMAR method with a perceptual loss. 
The RL-ARCNN~\cite{huang2018metal} introduced deep residual learning to reduce metal artifacts in cervical CT images and Wang~\etal~\cite{wang2018conditional} proposed to use the conditional generative adversarial network (cGAN)~\cite{isola2017image} to reduce metal artifacts in CT images. 
Very recently, Lin~\etal~\cite{lin2019dudonet} developed a dual-domain learning method to improve the image-restoration-based MAR results by involving sinogram enhancement as a procedure.
These image-restoration-based methods demonstrated good performance on their experimental datasets due to the powerful representation capability of deep neural networks.
However, in our experiments, we find that these methods tend to degrade on other site data, as the training samples hardly cover the unseen artifacts patterns.
\revise{Although DuDoNet~\cite{lin2019dudonet} introduces the sinogram enhancement procedure to improve the network performance, it still directly adopts the image-domain-refinement output (CNN output) as the final reconstructed image.}
As there is no geometry (physical) constraints to regularize the neural networks, there would be some tiny anatomical structure changes in the output image (see Fig.~\ref{fig:visualreal} for an example), which limits the usage of image domain methods in real clinical scenarios.

% our motivation
In this work, we present a novel image and sinogram domain joint learning framework for generalizable metal artifact reduction.
Different from the previous image-restoration-based solutions, we formulate the MAR as the deep-learning-based sinogram completion task and train a deep neural network, \ie, SinoNet, to restore the unreliable projections within the metal trace region.
To ease the SinoNet learning and improve the completion quality, we simultaneously train another neural network, \ie, PriorNet, to generate a good prior image with less metal artifact and guide the SinoNet learning with the forward projection of the prior image; see a sinogram completion result in Fig.~\ref{fig:sinocompletion}.
Moreover, we  design a novel residual sinogram learning strategy to fully utilize the prior sinogram guidance to improve the continuity of sinogram completion and thus alleviate the new artifacts in the reconstructed CT images.
The final CT image is then reconstructed from the completed sinogram with the conventional FBP algorithm.
Compared with the previous prior-image-based MAR approaches, the whole framework is trained in an efficient end-to-end manner so that the prior image generation and deep sinogram completion procedures can be learned in a collaborative manner and benefit from each other.
We extensively evaluate our framework on \revise{CT images with simulated and real metal artifacts}, demonstrating that our method produces superior artifact-reduced results and outperforms other MAR methods.

Our main contributions are summarized as follows.
\begin{enumerate}
	\item We present a novel image and sinogram domain joint learning framework for metal artifact reduction by simultaneously leveraging the advantages of image domain and sinogram domain-based MAR techniques. The proposed framework achieves superior performance on \revise{CT images with simulated and real metal artifacts}.

	\item We propose to train a deep prior image network to provide a good estimation of missing projections and thus enhance sinogram completion network learning. The two networks are trained in an end-to-end manner and can benefit from each other.

	\item We design a novel residual sinogram learning scheme to facilitate sinogram completion. The scheme is able to fully utilize prior image information and alleviate the new artifacts on the reconstructed CT image.
	
\end{enumerate}

The remainders of this paper are organized as follows.
We elaborate our framework in Section~\ref{sec:method}.
The experiments and results are presented in Section~\ref{sec:experiment}.
We further discuss the key issues of our method in Section~\ref{sec:discussion} and draw the conclusions in Section~\ref{sec:conclusion}.

	\begin{figure*}[t]
	\centering
	\includegraphics[width=1.0\linewidth]{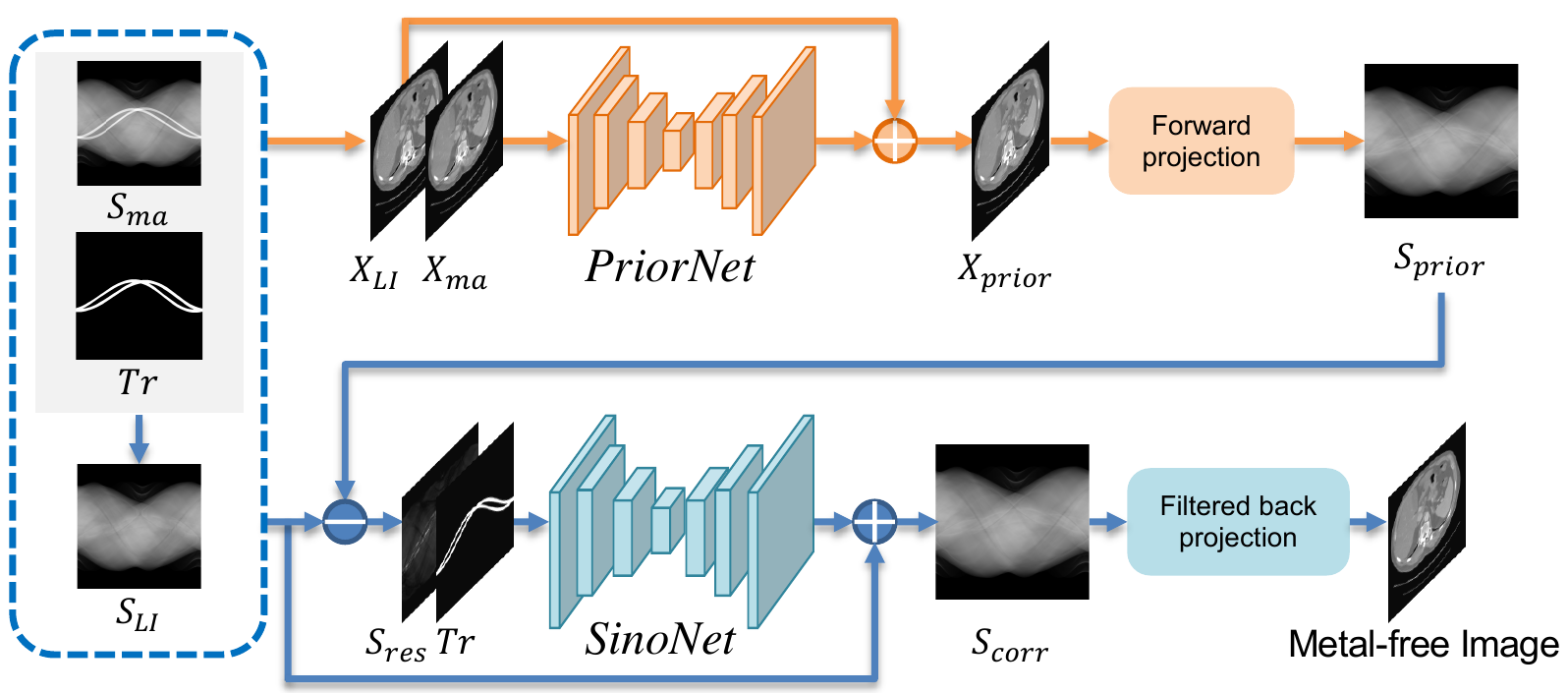}
	\caption{\revise{Schematic diagram of our proposed image and sinogram domain joint learning framework for metal artifact reduction.} Given the metal-affected sinogram $S_{ma}$ and metal trace mask $Tr$, we  use linear interpolation to acquire LI corrected sinogram $S_{LI}$. 
	We jointly train a prior image generation network, \ie, PriorNet, to generate a good prior image $X_{prior}$ and a sinogram completion network, \ie, SinoNet, to restore the metal-affected sinogram with the guidance of the prior sinogram $S_{prior}$, which is the forward projection of the prior image $X_{prior}$. \revise{The $S_{res}$ is the residual sinogram map between $S_{LI}$ and $S_{prior}$.} The final metal-free image is reconstructed from the corrected sinogram $S_{corr}$ with the FBP algorithm. }
	\label{fig:network_arch}
	\centering
	\vspace{-0.5cm}
\end{figure*}

\section{Methodology}
\label{sec:method}

\subsection{Overview}
Fig.~\ref{fig:network_arch} depicts the overview of our proposed image and sinogram domain joint learning framework for metal artifact reduction in CT images.
The whole framework integrates the image domain learning (prior image generation) and sinogram domain learning (sinogram completion).
Given the original metal-corrupted sinogram $S_{ma} \in \mathbb{R}^{H\times W}$ and the metal trace mask $Tr \in \{0,1\}^{H\times W}$, we first apply the linear interpolation~\cite{kalender1987reduction} to produce an initial estimation for the projections within the metal trace region and acquire the LI corrected sinogram $S_{LI}$ for the following procedures. 
To ease the sinogram completion procedure, we train an image domain network, \ie, PriorNet, to produce a good prior image $X_{prior}$ with less metal artifact and acquire the prior sinogram $S_{prior}$ with the forward projection of the generated prior image $X_{prior}$ to guide the sinogram domain learning.
We simultaneously train another deep neural network, \ie, SinoNet, to restore the metal-affected projections to acquire the corrected sinogram $S_{corr}$ by taking the LI corrected sinogram $S_{LI}$, the prior sinogram $S_{prior}$, and the metal trace mask $Tr$ as input. 
Particularly, we design a novel residual learning strategy and make the SinoNet refine the residual sinogram map between $S_{LI}$ and $S_{prior}$.
The final metal-free CT image is then reconstructed from the corrected sinogram $S_{corr}$ with the conventional FBP algorithm.
The whole framework is trained in an end-to-end manner so that the prior image generation and sinogram completion procedures can benefit from each other.

\subsection{Deep Prior Image Generation}
In this step, we propose to generate a prior image with a deep neural network to facilitate the sinogram completion procedure, as the metal-free prior image would provide a good estimation for the missing projections in the original sinogram.
A straightforward solution for this procedure is to take the original CT image with metal artifacts as input and train a neural network to generate the prior image with less metal artifact.
However, when the metal objects are relatively large, the metal artifacts in the original CT image would be very strong and it is difficult for the neural network to reduce the metal artifacts.
Therefore, besides the original CT image, we also involve the LI corrected image into the prior image generation procedure and employ a neural network, \ie, PriorNet, to refine the LI corrected image by residual learning.
Specifically, we first reconstruct the original metal-corrupted CT image $X_{ma}$ and LI corrected image $X_{LI}$ from the original metal-affected sinogram $S_{ma}$ and the linear interpolated sinogram $S_{LI}$, respectively.
Then the artifact-reduced prior image is represented as
\begin{equation}
X_{prior} = X_{LI} + f_{P} ([X_{ma}, X_{LI}]),
\end{equation}
where $f_{P}$ denotes the prior image generation network and $[a, b]$ represents the concatenation operation of image $a$ and $b$.

The PriorNet is based on the U-Net~\cite{ronneberger2015u} architecture, but we halve the channel number to reduce the total number of parameters.
To optimize the network, we  employ the L1 loss to minimize the difference between the network output and the  ground truth CT image $X_{gt}$ without metal artifacts
\begin{equation}
\mathcal{L}_{prior} = || X_{prior} - X_{gt}||_1.
\end{equation}
We further acquire the prior sinogram $S_{prior}$ by performing forward projection on the generated prior image $X_{prior}$
\begin{equation}
S_{prior} = \mathcal{P}(X_{prior}), 
\end{equation}
where $\mathcal{P}$  denotes the forward projection operator.
The prior sinogram $S_{prior}$ is then used to guide the network to complete the missing projections in the sinogram domain.

\subsection{Deep Sinogram Completion}
With the guidance of prior sinogram $S_{prior}$, we train another neural network, \ie, SinoNet, to restore the projections within the metal trace region $Tr$ in the sinogram domain.
Specifically, the SinoNet takes the LI corrected sinogram $S_{LI}$, the prior sinogram $S_{prior}$, and the metal trace $Tr$ as input, and outputs the missing projections in metal trace region $Tr$ by utilizing the contextual information of the sinogram.
To improve the continuity of the completed projections at the boundary of the metal trace region, we design a \textit{residual sinogram learning} strategy and make the SinoNet refine the residual sinogram between  $S_{LI}$ and $S_{prior}$. 
Particularly, we calculate the residual sinogram map $S_{res}$, which can be treated as a smooth transition between the prior sinogram $S_{prior}$ and the LI corrected sinogram $S_{LI}$, and then we employ the SinoNet to refine the \textit{residual projections} within the metal trace region $Tr$. The corrected sinogram $S'_{corr}$ can be written as:
 \begin{align}
 \centering
 S'_{corr} &= f_{S}([S_{prior} -S_{LI}, Tr]) +S_{LI},  
 \end{align}
where $f_{S}$ represents the sinogram completion network.
As the network estimates the residual values instead of the absolute projection values, it can alleviate the discontinuity at the boundary of the metal trace~\cite{zhang2013hybrid}.
Considering that the metals only affect projection data in metal trace region, we further composite the output of SinoNet and $S_{LI}$ with respect to $Tr$ to get the final corrected result:
\revise{
\begin{align}
	S_{corr} &= S'_{corr}\odot Tr + S_{LI} \odot (1-Tr)\\ \nonumber
	 			 &= f_{S}([S_{prior} -S_{LI}, Tr])\odot Tr + S_{LI},
\end{align}
}
where  $\odot$ denotes element-wise multiplication. 

To estimate the residual projections within the metal trace region, the network should be better aware of the metal trace information.
However, the metal mask or metal trace regions are usually small and occupy a small portion of the whole sinogram, directly concatenating the residual sinogram map $S_{prior}-S_{LI}$ and $Tr$ as network input would weaken the metal trace information due to the down-sampling operations of the network.
Therefore, we employ the mask pyramid U-Net~\cite{liao2019generative} to retain the metal trace information into each layer explicitly so that the network is able to extract more discriminative feature for restoring the missing information at metal trace region.

To optimize the SinoNet, we adopt the L1 loss to minimize the differences between the corrected sinogram and the ground truth sinogram $S_{gt}$ without metals.
However, as the composited sinogram $S_{corr}$ has the identical values with the ground truth sinogram outside the metal trace region, directly minimizing the difference between $S_{corr}$ and $S_{gt}$ would provide supervision for the network output only within the metal trace region.
As we mentioned above, the metal region occupies only a small portion of the whole sinogram.
To improve the training efficiency,  we also encourage the pre-composited  sinogram $S'_{corr}$ to be close to the ground truth sinogram $S_{gt}$ so that the loss function can also provide supervision for those network outputs outside of the metal trace region. 
The total objective of the deep sinogram completion can be represented as
\begin{equation}
\mathcal{L}_{sino} = || S_{gt} - S_{corr}||_1+\beta|| S_{gt} - S'_{corr}||_1,
\end{equation}
where $\beta$ is a hyper-parameter to control the trade-off between two difference items. we find that it is not sensitive to the network performance and we empirically set it as 0.1 in our experiments.

\subsection{Overall Objective Function and Technical Details}
The above L1 loss for SinoNet optimization only penalizes single projection value inconsistency in the sinogram domain, without considering the geometry-consistency of the completed values or penalizing the new artifacts in the reconstructed CT images.
Therefore, we  further design a  filtered back-prorogation (FBP)  loss to alleviate the new artifacts in the reconstructed CT image 
\revise{
\begin{equation}
\begin{aligned}
\mathcal{L}_{FBP} &=||(\mathcal{P}^{-1}(S_{corr})- X_{gt})\odot (1-M)||_1,
\end{aligned}
\end{equation}
}
where $\mathcal{P}^{-1}$ represents the FBP operator and $M$ is the metal mask.
Here we adopt the masked L1 loss to penalize the intensity difference only in the non-metal regions, as it is difficult to accurately reconstruct the original image at the metal position. 
Note that the FBP operation $\mathcal{P}^{-1}$ is \textit{differentiable}, so that the gradient of $\mathcal{L}_{FBP}$ is able to back-propagate to SinoNet, encouraging it to generate geometry-consistent completion results.
We jointly train the PriorNet and SinoNet in an end-to-end manner and the total objective function is
\begin{equation}
\mathcal{L} _{total}= \mathcal{L} _{prior} + \alpha_{1}\mathcal{L} _{sino}+ \alpha_{2}\revise{\mathcal{L} _{FBP}},
\end{equation}
where $\alpha_{1}$ and $\alpha_{2}$ are hyperparameters to balance the weight of different loss items. 
We empirically set them as 1.0 in our experiments.

Our whole framework takes original metal-affected sinogram and metal trace as input. 
In the training phase,  we use the simulated data to train the whole framework so that we can acquire the metal trace mask $Tr$ for the simulated training data by performing the forward projection on the simulated metal mask $M$.
In the testing phase, given the metal-affected sinogram $S_{ma}$, we can segment the metal mask $M$ from the reconstructed metal-corrupted CT images $X_{ma}$ with simple thresholding method or other advanced metal segmentation algorithms, and then conduct the similar forward projection to get the metal trace mask $Tr$.

	\section{Experiments}
\label{sec:experiment}

\begin{figure*}[!htbp]
	\centering
	\includegraphics[width=0.85\textwidth]{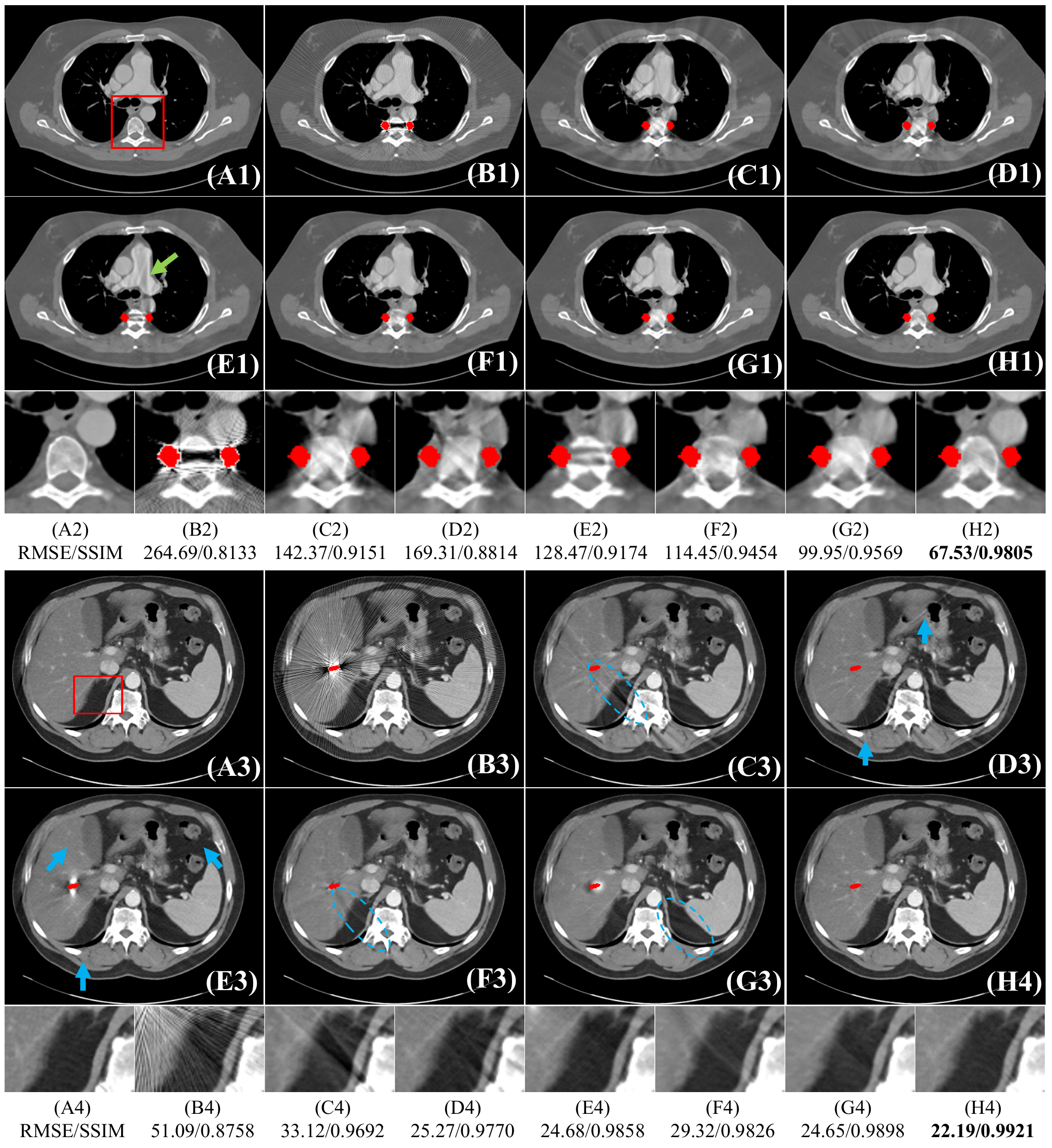}
	\caption{\revise{Visual comparison with different methods on DeepLesion dataset. The simulated metal masks are colored in red for better visualization. The (A1-A4) are refer images. We show the MAR results of LI~\cite{kalender1987reduction} (C1-C4), NMAR~\cite{meyer2010normalized} (D1-D4), CNNMAR~\cite{zhang2018convolutional} (E1-E4), cGANMAR~\cite{wang2018conditional} (F1-F4), DuDoNet~\cite{lin2019dudonet} (G1-G4), and our method (H1-H4). The display window of the first and second samples are [-480 560] and [-175, 275] HU, respectively. We also use ROI RMSE and ROI SSIM to show quantitative results for a better comparison.}} 
	\label{fig:visualdeeplesion}
\end{figure*}

\begin{figure*}[!htbp]
	\centering
	\includegraphics[width=0.9\textwidth]{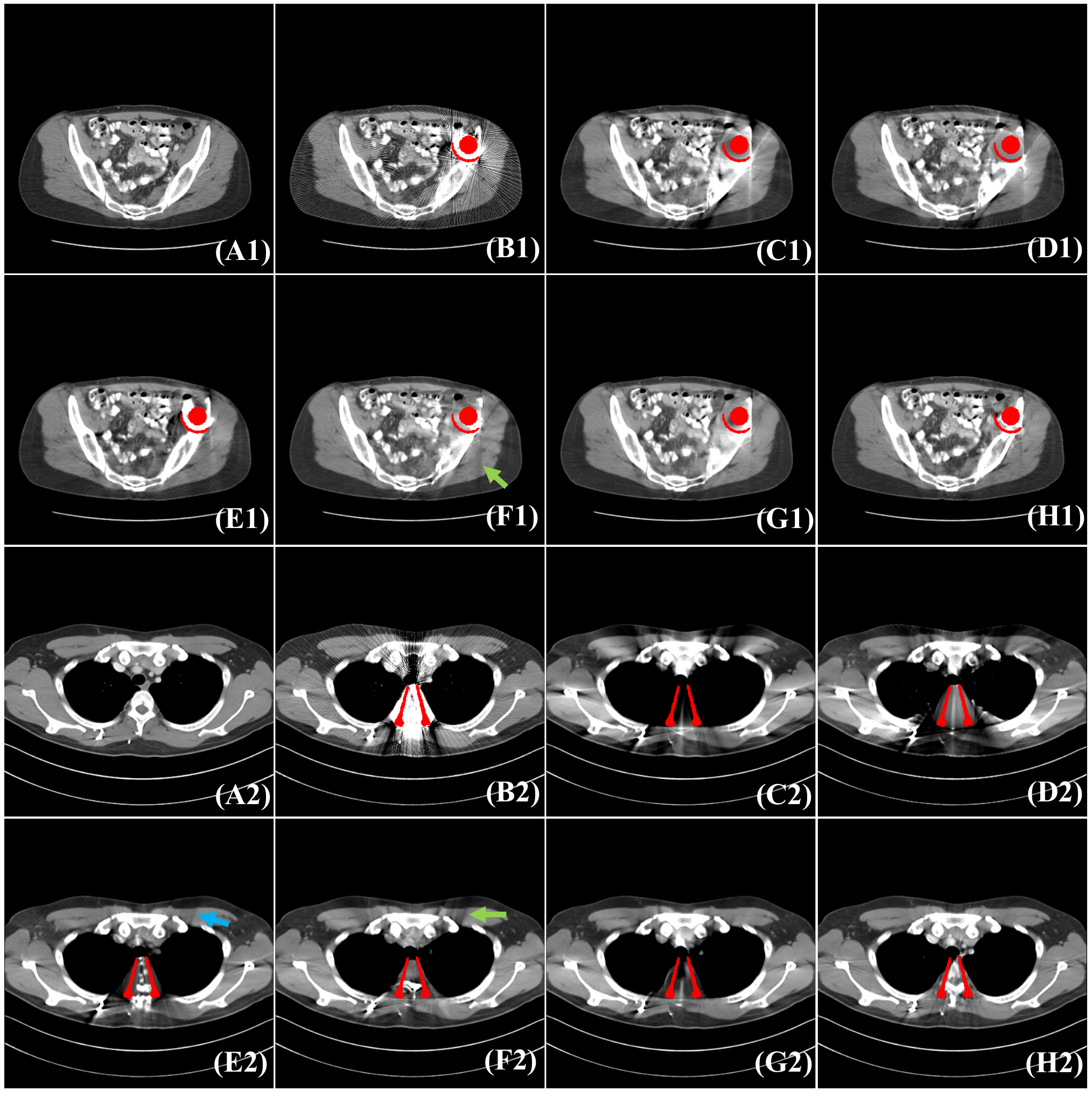}
	\caption{\revise{Visual comparison with different methods on DeepLesion dataset. The simulated metal masks are colored in red for better visualization.
	The (A1-A2) are refer images. We show the MAR results of LI~\cite{kalender1987reduction} (C1-C2), NMAR~\cite{meyer2010normalized} (D1-D2), CNNMAR~\cite{zhang2018convolutional} (E1-E2), cGANMAR~\cite{wang2018conditional} (F1-F2), DuDoNet~\cite{lin2019dudonet} (G1-G2), and our method (H1-H2). The display window is [-175 275] HU.}} 
	\label{fig:visualdeeplesion2}
\end{figure*}

\subsection{Dataset and  Simulation}
We evaluated our method with simulated metal artifacts on CT images and real CT images with metal artifacts.
For the simulation data, we randomly selected a subset of CT images from the recently released DeepLesion dataset~\cite{yan2018deep} to synthesize metal artifacts.
For the simulated metal masks, we employed the previous metal mask collection in~\cite{zhang2018convolutional}, which contains 100 manually segmented metal implants with different shapes and sizes.
Specifically, we randomly chose 1000 CT images and 90 metal masks to synthesize the training data. 
The remaining 10 metal masks were paired with an additional 200 CT images from 12 patients to generate 2000 combinations for network evaluation.

We followed the procedure in~\cite{zhang2018convolutional,lin2019dudonet} to simulate the metal-corrupted sinograms and CT images by inserting metallic implants into clean CT images, \revise{where beam hardening and Poisson noise are simulated.}
We employed a polychromatic X-ray source and assumed the incident X-ray has $2\times10^7$ photons. 
The partial volume effect was also considered during the simulation.
A fan-beam geometry was adopted and we uniformly sampled 640 projection views between 0-360 degrees.
Before the simulation, the CT images were resized to $416\times416$, resulting in the sinogram with the size of $641\times640$.

\subsection{Implementation Details}
The framework was implemented in Python based on PyTorch~\cite{paszke2019pytorch} deep learning library.
We trained the PriorNet and SinoNet in an end-to-end manner with differential forward projection (FP) and filtered backprojection (FBP)  operations \revise{provided in ODL library\footnote{\revise{https://github.com/odlgroup/odl}}.}
In the network training, all the images had a size of $416\times416$ and the sinograms were with a size of $641\times640$.
The Adam optimizer~\cite{kingma2014adam} was used to optimize the whole framework with the parameters $(\beta1, \beta2)=(0.5, 0.999)$.
We totally trained 400 epochs with a mini-batch size of 8 on one Nvidia 1080Ti GPU and the learning rate was set as $1e^{-4}$.
In each training iteration, we randomly chose one CT image with synthesized metal artifacts from the pool of 90 different metal mask pairs and the different CT images were formed as one mini-batch data to be fed into the network for calculating the total objective function.

\subsection{Experimental Results on DeepLesion Data}

\begin{table} [t]
	\centering
	\caption{\revise{Quantitative comparison of different methods on DeepLesion dataset.}}
	\label{table:deeplesion}
	\begin{tabular}{p{3cm}<{\centering}|p{2cm}<{\centering}|p{2cm}<{\centering}}
		\toprule[1pt]
		Method & RMSE (HU) & SSIM\\ \hline
		
		LI~\cite{kalender1987reduction}					 &50.31$\pm$19.41       &0.9455$\pm$0.0315 \\  	
		NMAR~\cite{meyer2010normalized}				&47.03$\pm$20.67     &0.9594$\pm$0.0299 \\  
		CNNMAR~\cite{zhang2018convolutional}	&43.27$\pm$14.44     &0.9706$\pm$0.0159 \\ 	
		cGANMAR~\cite{wang2018conditional}		 & 39.01$\pm$12.66	  &0.9754$\pm$0.0055  \\ 
		DuDoNet~\cite{lin2019dudonet}     &38.00$\pm$13.31      &0.9766$\pm$0.0072   \\ \hline
		\textbf{Ours}      &\textbf{31.15$\pm$5.81}    &\textbf{0.9784$\pm$0.0048}		\\	
		\bottomrule[1pt]
	\end{tabular}
\end{table}

\begin{figure*}[t]
	\centering
	\includegraphics[width=1.0\textwidth]{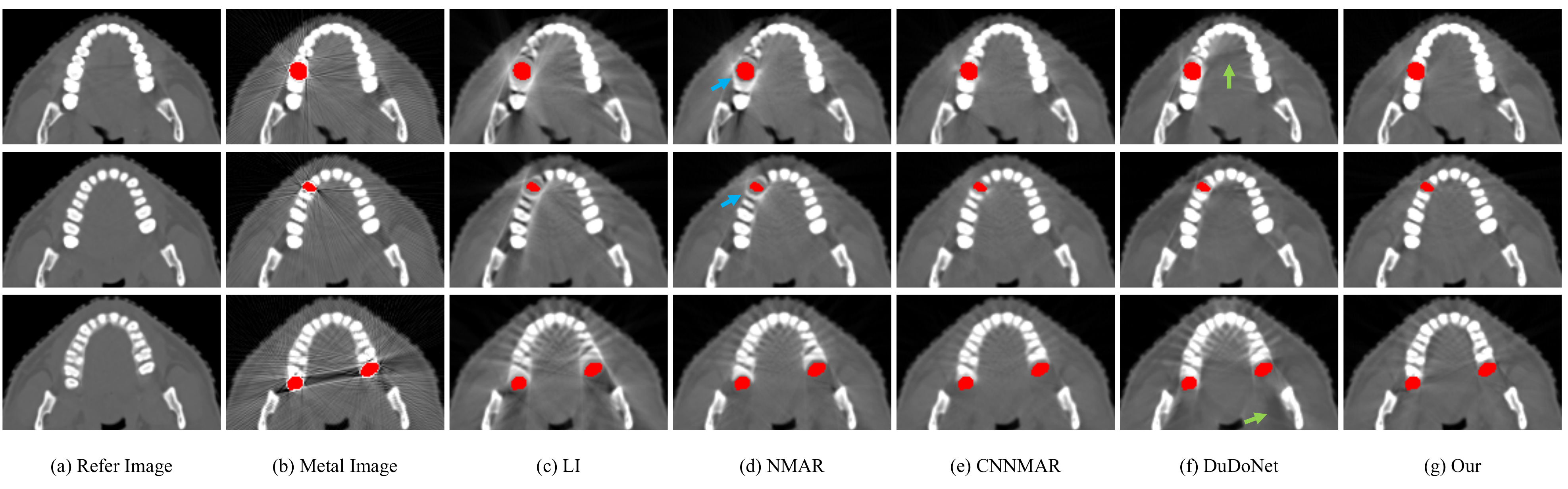}
	\caption{\revise{Visual results on head CT images with different numbers of simulated dental fillings. The simulated  dental fillings are colored in red for better visualization. The display window is [-1000 1600] HU.}} 
	\label{fig:visualhead}
\end{figure*}

\subsubsection{Quantitative comparisons with state-of-the-art methods}
We compare our method with conventional interpolation-based methods: linear interpolation (LI)~\cite{kalender1987reduction} and normalized metal artifact reduction (NMAR)~\cite{meyer2010normalized}, which are widely employed approaches in MAR.
Also, we compare our method with the recent deep-learning-based methods CNNMAR~\cite{zhang2018convolutional}, cGANMAR~\cite{wang2018conditional}, and the state-of-the-art method DuDoNet~\cite{lin2019dudonet}.
The CNNMAR approach also adopts  a CNN to generate a reduced-artifact prior image and then uses traditional interpolation to correct the metal-corrupted regions in the sinogram.
\revise{Both cGANMAR and DuDoNet employ the image domain network to generate the final results, where cGANMAR directly uses an image-to-image translation network to reduce artifacts on original metal images while DuDoNet further incorporates sinogram enhancement to ease image domain learning.}
For CNNMAR, we used the public released code and model.
We re-implemented DuDoNet~\cite{lin2019dudonet} and cGANMAR~\cite{wang2018conditional}, since there is no public implementation.

Table~\ref{table:deeplesion} shows the quantitative comparison results of our method and other methods on the DeepLesion dataset.
It is observed that the prior-image-based interpolation method NMAR outperforms LI approach on both root mean square error (RMSE) and structured similarity index (SSIM) metrics, as the prior image information improves the accuracy of interpolation for missing projection values.
The deep-learning-based methods CNNMAR and cGANMAR achieves much lower RMSE and higher SSIM values than conventional MAR methods, showing the advantage of data-driven deep neural networks for metal artifact reduction.
\revise{The DuDoNet achieves better RMSE and SSIM performance when compared with cGANMAR}, as it integrates sinogram enhancement to reduce artifacts before conducting the image refinement procedure.
\revise{Compared with DuDoNet, our method further reduces RMSE with 6.85 HU and achieves slightly better SSIM values.} 
Overall, our framework attains the best performance among different methods in terms of RMSE and SSIM, showing the effectiveness of our method for metal artifact reduction.

\subsubsection{Qualitative analysis}
Fig.~\ref{fig:visualdeeplesion} and Fig.~\ref{fig:visualdeeplesion2} shows the visual comparisons of our method and other methods on DeepLesion simulation data.
We show the refer metal-free CT images, simulated metal artifact images (Metal Image), and metal artifact reduction results of different methods.
The simulated metal masks are colored in red for better visualization.
It is observed that severe streaking artifacts are in the original metal images and a severe dark strip exists between two metal implants (\revise{see Fig.~\ref{fig:visualdeeplesion}(B1))}. 
Generally, the deep-learning-based methods CNNMAR, \revise{cGANMAR} and DuDoNet can reduce more artifacts than conventional LI and NMAR approaches.
When the metal implants are small (Fig.~\ref{fig:visualdeeplesion}), \revise{the DuDoNet and cGANMAR can achieve better visual results than CNNMAR, while there are still some mild artifacts in the DuDoNet and cGANMAR results; see the dashed blue ovals in Fig.~\ref{fig:visualdeeplesion}(F3\&G3).}
Compared with these methods, our method effectively reduces most artifacts and retains the fine details of the structures.
\revise{Fig.~\ref{fig:visualdeeplesion2} shows the results when the metal implants are large.
It is observed that the conventional interpolation methods LI and NMAR, and image domain method cGANMAR and DuDoNet cannot preserve the details of the original structures and there are some new secondary-artifacts in the cGANMAR results; see the green arrows in Fig.~\ref{fig:visualdeeplesion2}(F1\&F2).
}
Compared to CNNMAR, our method preserves better structure details, showing the effectiveness of the deep signogram completion mechanism. 
\revise{We also calculate the ROI RMSE and ROI SSIM for the red box patches in Fig.~\ref{fig:visualhead} to quantitatively compare different methods. As shown in Fig.~\ref{fig:visualhead} (A2-H2\&A4-H4), our method achieves the lowest RMSE and highest SSIM values among different methods.}

\begin{figure*}[!htbp]
	\centering
	\includegraphics[width=0.85\textwidth]{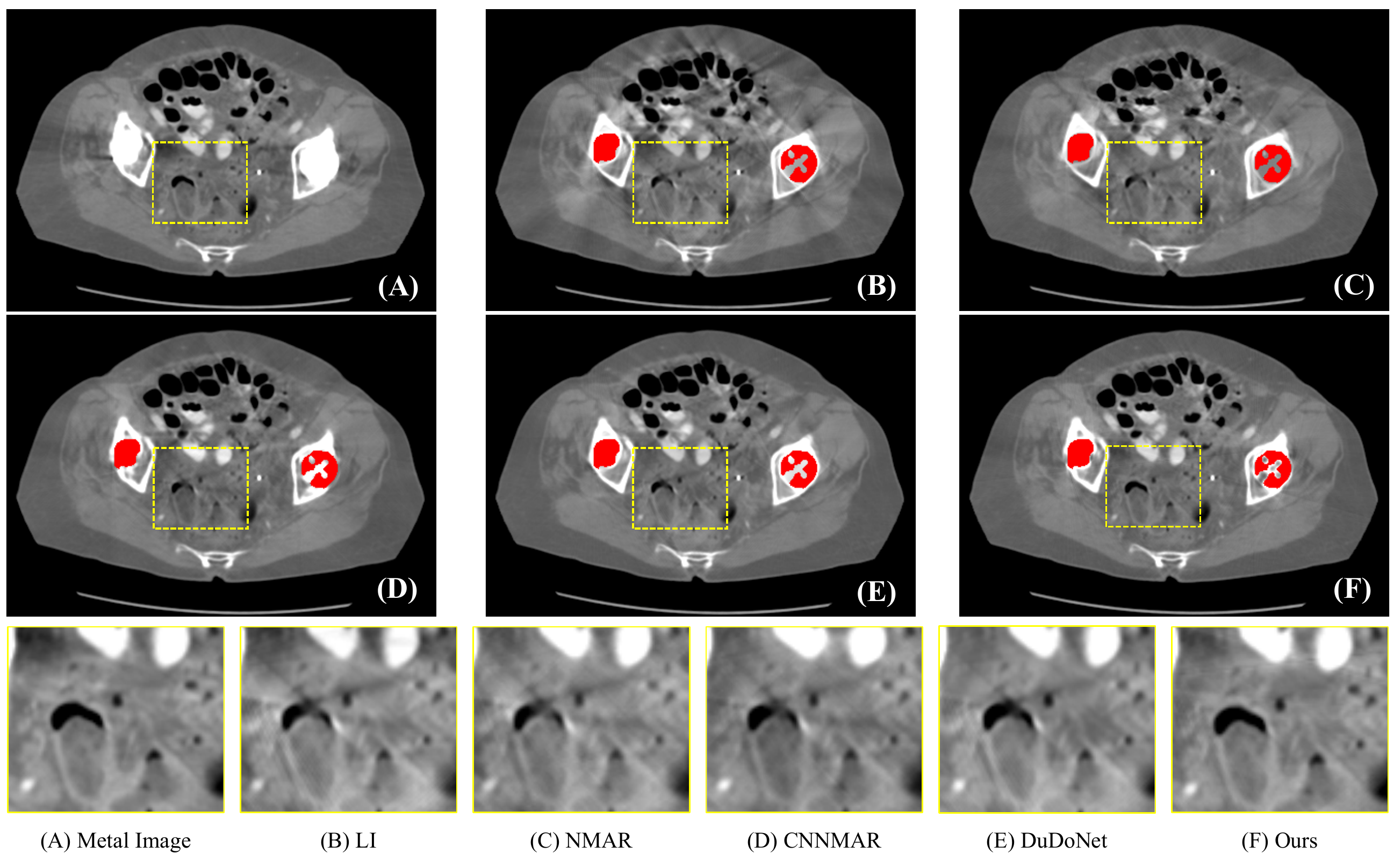}
	\caption{\revise{Visual comparison on CT images with real metal artifacts. The segmented metals are colored in red for better visualization. Our method effectively reduces metal artifacts and preserves the fine-grained anatomical structures. The display window of whole image is [-480 560] HU and the display window of cropped patches is [-400 300] HU.}}
	\label{fig:visualreal}
\end{figure*}

\subsection{Generalization to Different Site Data}
The selected CT images from the DeepLesion dataset are samples of abdomen and thorax.
To show the feasibility of our method applied to different site data, we directly evaluate the model trained with DeepLesion data on the head CT images \revise{collected from the online website} with simulated metal artifacts. 
Fig.~\ref{fig:visualhead} shows the visual metal artifact reduction results of our method on the head CT images with simulated dental fillings.
\revise{We also show the conventional LI and NMAR results and the deep-learning-based CNNMAR and DuDoNet results.}
\revise{It is observed that the LI and NMAR introduce several secondary-artifacts and could change the anatomical structures of the tooth; see blue arrows in Fig.~\ref{fig:visualhead}(d). The DuDoNet can further reduce the artifacts, while there are still several shading artifacts in the output images; see green arrows in Fig.~\ref{fig:visualhead}(f). Although without training with head CT images, our method effectively reduces artifacts,indicating that the proposed method has the potential to handle different site data. Notably, the MAR result of our method is even comparable with CNNMAR, which is also trained with head CT images.}

\begin{figure}[!htbp]
	\centering
	\includegraphics[width=0.5\textwidth]{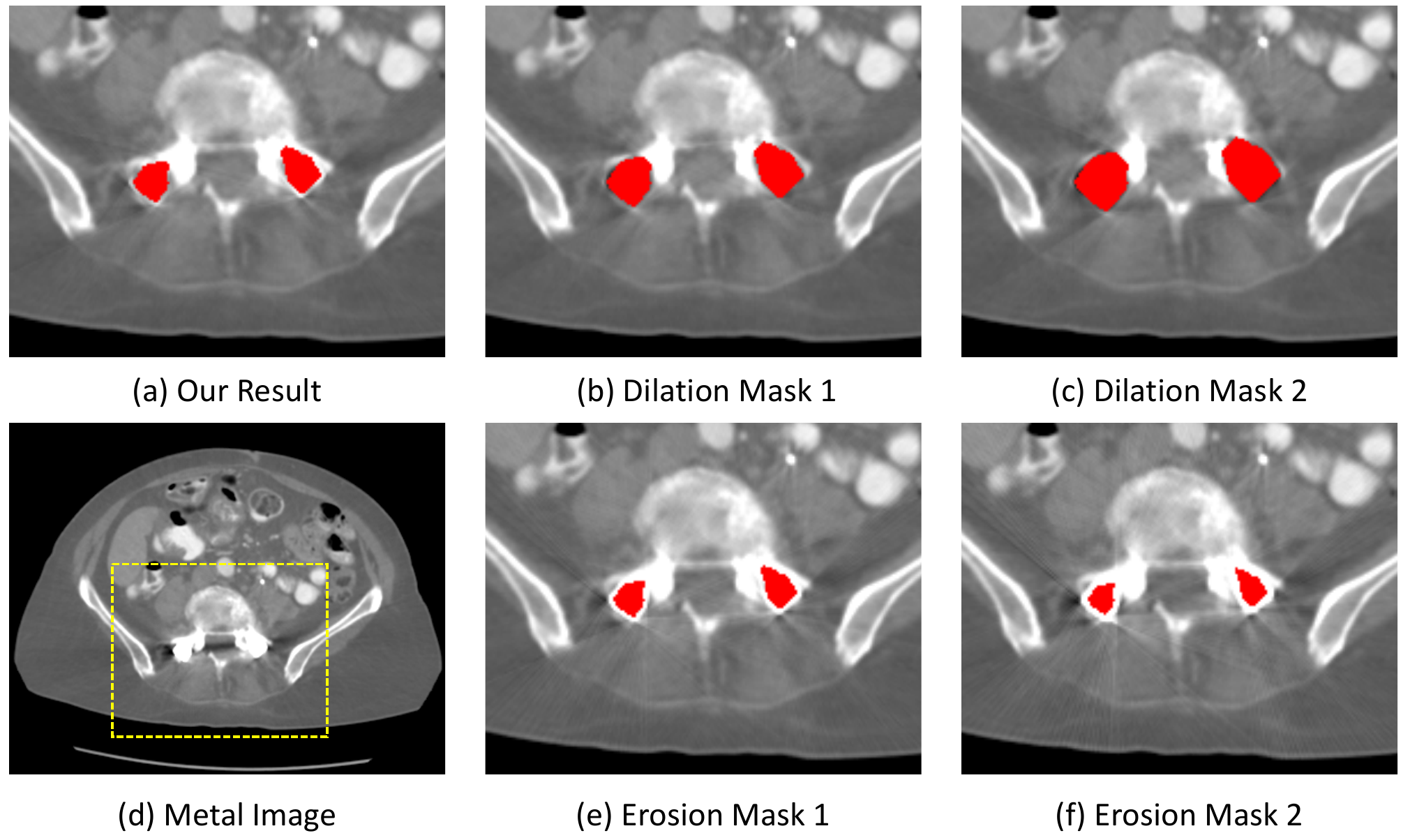}
	\caption{The results of our method with different segmented metal masks as input.  We show the MAR results with thresholding-based metal segmentation (a); we also dilate the metal mask to get over-segmented masks (b \& c) and erode the metal mask to get under-segmented masks (e \& f). The display window is [-480 560] HU.} 
	\label{fig:differentmask}
\end{figure}

\subsection{\revise{Experiments on CT Images with Real Metal Artifacts} }

\subsubsection{\revise{Results on  CT images with real metal artifacts}}
Since the original sinogram data with metal artifacts in the real clinical scenario are difficult to access, we follow the previous work~\cite{lin2019dudonet} to evaluate our method on clinical CT images with metal artifacts.
Specifically, we collected some clinical CT images with metal artifacts and segmented the metal mask from the clinical CT images with a simple thresholding method (\ie, 2000 HU in our experiments). 
The forward projection of the metal masks was conducted to generate the metal projection and the pixels with the projection value greater than zero were regarded as the metal trace region $Tr$.
We also performed the forward projection on the clinical CT image with the same imaging geometry as the above simulation procedure to acquire the metal-corrupted sinogram $S_{ma}$. 
The LI corrected sinogram $S_{LI}$ was then generated from $S_{ma}$ and $Tr$ with linear interpolation.
Finally, we fed $S_{ma}$, $S_{LI}$ and $Tr$ into our framework to get the meta artifact reduction images.
\revise{Fig.~\ref{fig:visualreal} presents the visual results of different methods. 
Our method effectively reduces metal artifacts compared with the original metal image. 
From the yellow zoomed patches, it is observed that the other methods change some tiny anatomical structures of the original image, while our method can preserve the fine-grained anatomical structures.}

\subsubsection{The influence of  metal mask segmentation}
An accurate metal trace mask (or metal mask equally) is vital for the good performance of metal artifacts reduction in our framework. 
In practice, we can manually segment the metals or use some automatic metal segmentation methods (\eg, thresholding method) to segment the metals.  
To investigate the influence of metal mask segmentation on the final MAR results, we take different metal masks to acquire the metal traces and then adopt our method to conduct MAR with taking these different metal traces as input.
Fig.~\ref{fig:differentmask} shows the MAR results of our method by taking the original thresholding-based metal mask (a), the dilation metal masks (b \& c), and erosion metal masks (e \& f) as input.
\revise{In general, our method can achieve reasonable MAR results under slightly metal segmentation errors.}
The over-segmented metal masks would lead to a relatively large metal trace and our method needs to complete more projection values.
As the network has been trained to completed different metal traces, it can estimate the missing projection values without leading to strong new artifacts in the reconstructed CT images, \revise{although there are additional shading artifacts in the CT images.}
\revise{As for the under-segmented case}, the corresponding metal trace is narrower than the original metal trace. The sinogram completion network would only complete some projection values of the original metal trace region and reuse some unreliable projection data. Therefore, our method can only reduce some metal artifacts \revise{and there still remains some residual streaking artifacts} in the reconstructed CT image.

\subsection{Analytical  Studies}

\begin{table} [t]
	\centering
	\caption{\revise{Quantitative analysis of our methods on simulated DeepLesion dataset.}}
	\label{table:ablationstudy}
	\begin{tabular}{p{3.6cm}<{\centering}|p{1.7cm}<{\centering}|p{2.0cm}<{\centering}}
		\toprule[1pt]
		Method & RMSE (HU) & SSIM\\ \hline
		
		Deep sinogram completion	              &43.65$\pm$17.61   & 0.9720$\pm$0.0082\\ 	
		With tissue processing  &35.64$\pm$7.91     &0.9768$\pm$0.0047   \\ 
		w/o residual sinogram learning	& 31.86$\pm$4.69	  &0.9781$\pm$0.0048    \\ 
		Only metal image           &31.60$\pm$4.95      &0.9778$\pm$0.0048\\ \hline
		\textbf{Ours}      &\textbf{31.15$\pm$5.81}    &\textbf{0.9784$\pm$0.0048}		\\	
		\bottomrule[1pt]
	\end{tabular}
\end{table}

\begin{figure}[t]
	\centering
	\includegraphics[width=0.48\textwidth]{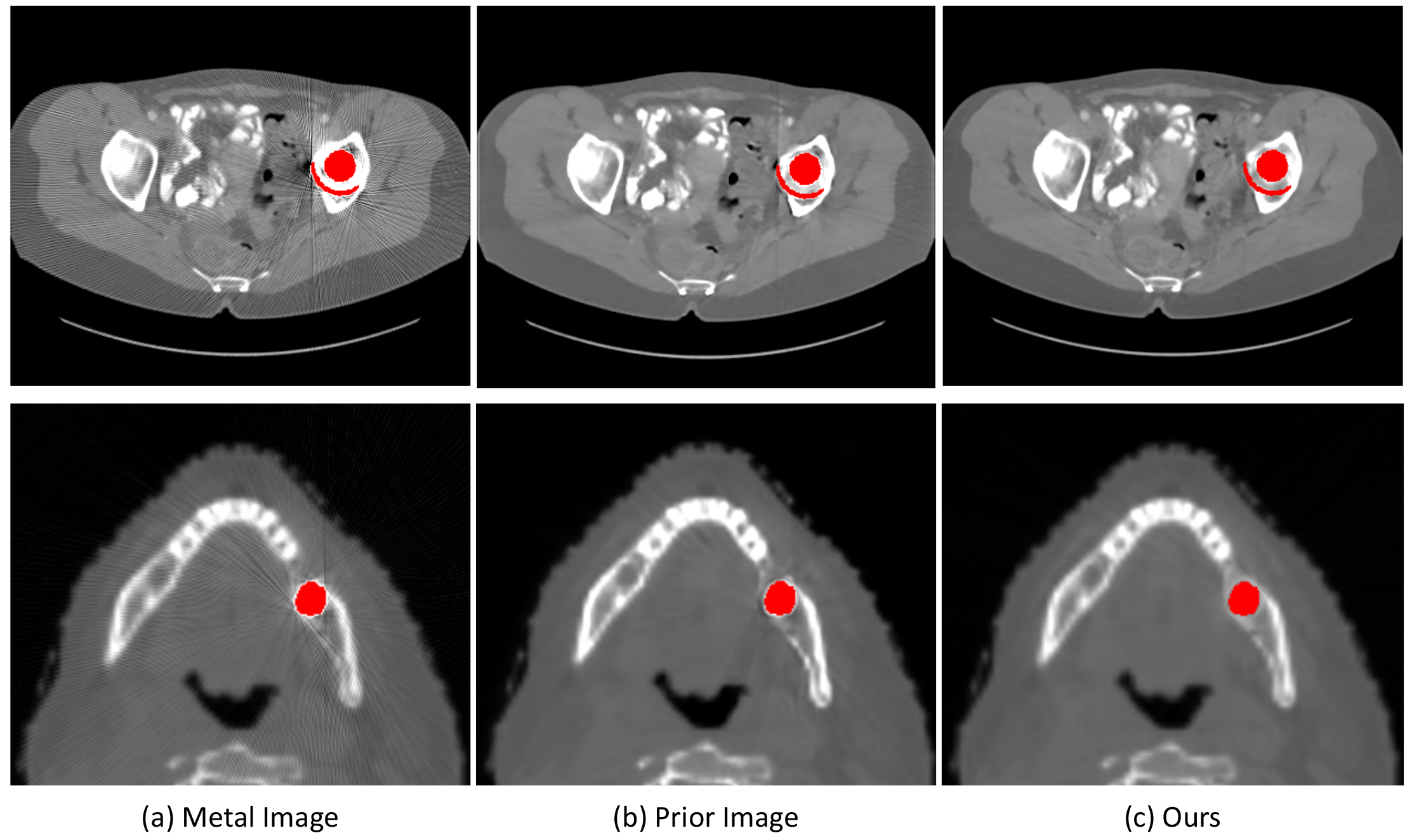}
	\caption{Illustration of the generated prior image. The display window of first row and seconed row are [-480 560] HU and  [-1000 1600] HU, respectively.} 
	\label{fig:priorimage}
\end{figure}

\subsubsection{Effectiveness of prior image generation}
In our framework, we train the PriorNet to generate a good prior image to ease sinogram completion. To show the effectiveness of this procedure, we directly train a neural network to complete the sinogram without taking the prior sinogram as input. The quantitative results of this method on DeepLesion simulated dataset are shown in Table~\ref{table:ablationstudy}.
It is observed that this method (Deep sinogram completion) achieves much higher RMSE and lower SSIM values than our method, verifying the effectiveness of prior image generation procedure.
In Fig.~\ref{fig:priorimage}, we show some generated prior images and final metal artifact reduction images on DeepLesion and head CT data. 
We can see that the prior images (Fig.~\ref{fig:priorimage}(b)) have less artifacts than original metal image (Fig.~\ref{fig:priorimage}(a)) and our final results (Fig.~\ref{fig:priorimage}(c)) further reduce artifacts compared with the prior images.
We also train our framework by taking only original metal image as PriorNet input.
As shown in Table~\ref{table:ablationstudy}, this method (Only metal image) generates slightly worse results than our method on the simulated dataset, indicating that incorporating LI corrected image as input can facilitate the prior image generation.

\subsubsection{Effectiveness of residual sinogram learning}
\begin{figure}[!tbp]
	\centering
	\includegraphics[width=0.48\textwidth]{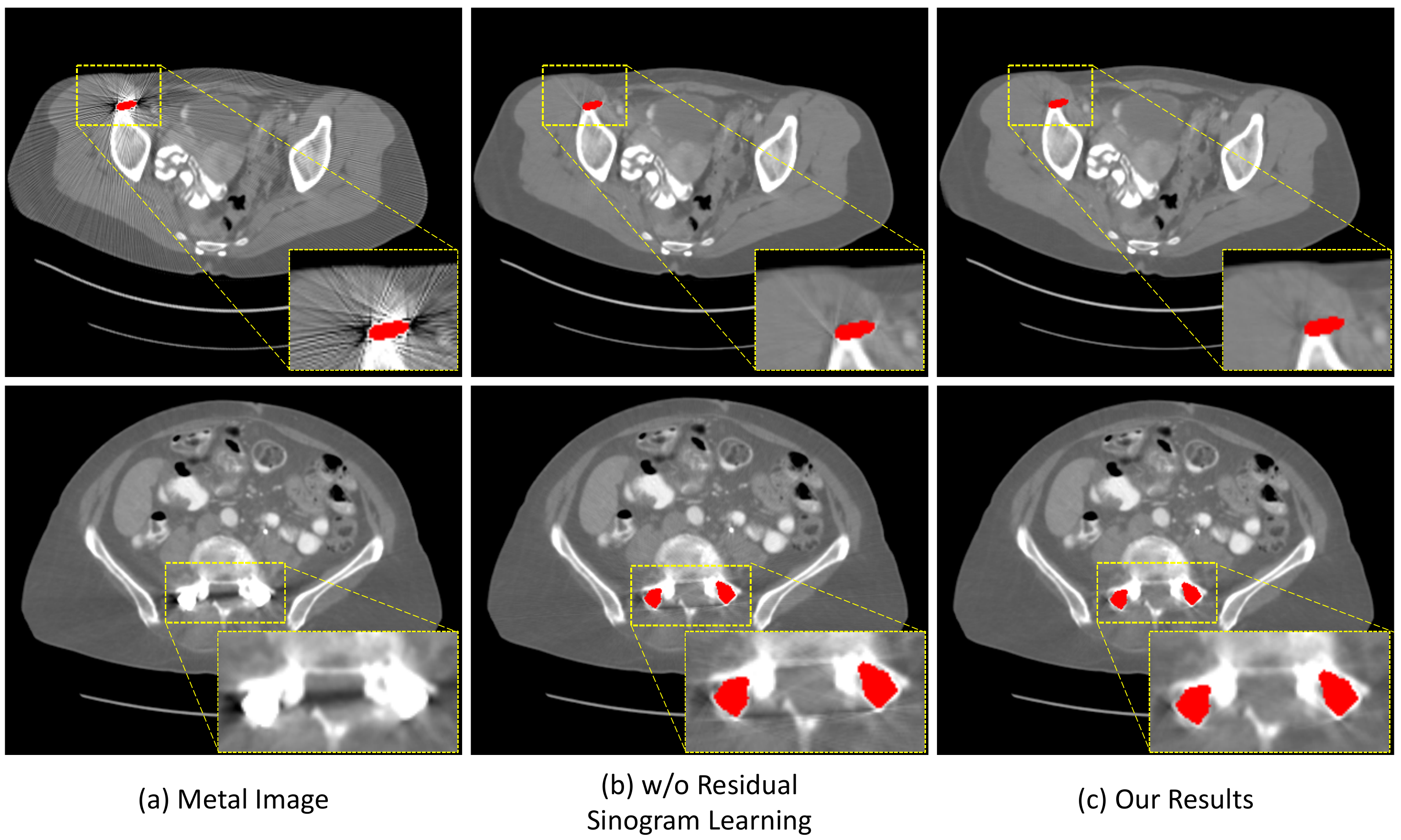}
	\caption{The results of our method with (b) and without (c) residual sinogram learning strategy. The first row is simulated metal image and the second image is real clinical CT image. The display window is [-480 560] HU.} 
	\label{fig:residuallearning}
\end{figure}

We show the qualitative MAR results of our method with and without residual sinogram learning strategy in Fig.~\ref{fig:residuallearning}. The first row is the results on the simulated metal artifact image and the second row shows the results on the real clinical CT image with metal artifacts.
In the experiment without residual sinogram learning, the SinoNet directly takes the prior sinogram and metal trace mask as input and outputs the refined projections within the metal trace region.
From the visual comparison in Fig.~\ref{fig:residuallearning}, we can observe that our framework further reduces metal artifacts on both simulated and real samples by adopting the residual sinogram learning strategy.
\revise{We also present the quantitative results of “w/o residual sinogram learning” in Table~\ref{table:ablationstudy}. It is observed that we achieve higher RMSE and lower SSIM values with residual sinogram learning. }

\subsubsection{Compared with tissue processing}
The tissue processing step is often used to acquire the prior image in previous MAR methods.
We conduct another experiment to investigate the effect of this strategy, where we employ the tissue processing and metal trace replacement steps~\cite{zhang2013hybrid, zhang2018sparse} to process the generated prior image $X_{prior}$ of our method and then acquire the final metal-free image with FBP reconstruction.
The quantitative result of this method (With tissue processing) is shown in Table~\ref{table:ablationstudy}. 
It is observed that \emph{With tissue processing} achieves satisfying results on the simulated DeepLesion dataset, while its performance is still inferior to our end-to-end deep sinogram completion strategy, indicating that the deep sinogram network could automatically learn how to reduce the mild artifacts in the prior image.

	\section{DISCUSSION}
\label{sec:discussion}

MAR is a long-standing problem in CT imaging. 
In this work, we aim to design a data-driven framework to address this problem by utilizing a large amount of training data.
The previous deep-learning-based methods usually formulate the MAR as an image restoration problem.
Whereas we borrow the spirit of conventional MAR approaches and formulate the MAR as a deep sinogram completion problem,  aiming to improve the generalization and robustness of the framework.  
Since directly regressing the accurate missing projection data is difficult, we propose to incorporate the deep prior image generation procedure and adopt a residual sinogram completion strategy.
This manner can improve the continuity of the projection values at the boundary of metal traces and alleviate the new artifacts, which are the common drawbacks of sinogram completion based MAR methods. 
In such way, our framework could better utilize the advantages of deep learning techniques while alleviating the risk of overfitting to certain training data.

\revise{We solve MAR in both sinogram and image domains, which share the same strategy with  DuDoNet~\cite{lin2019dudonet} and DuDoNet++~\cite{lyu2020dudonet}.
However, our framework differs from them in a few important aspects. 
The DuDoNet and DuDoNet++ directly adopt the image-domain-refinement output as the final MAR image, whereas the final MAR image of our method is directly FBP-reconstructed from the completed sinogram. As there is no geometry (physical) constraints to regularize the neural networks, there would be some tiny anatomical structure changes in the CNN-output images. 
Our FBP-reconstructed image could preserve the anatomical structure of the original image and avoid the resolution loss, as we only modify the metal trace region values in the sinogram; see comparisons in Figs.~\ref{fig:visualdeeplesion}\&\ref{fig:visualdeeplesion2}. 
More importantly, we design a novel residual learning strategy for the sinogram enhancement network to refine the residual projections within the metal trace region, and both quantitative and qualitative results show the effectiveness of such residual learning strategy.}

It is clinically impractical to acquire metal-free and metal-inserted CT data for network training, 
We thus simulate metal artifacts from clinical metal-free CT images to acquire synthesized training pairs.
In this case, the quality of simulated data would largely influence network performance.
Currently, we simulate the metal artifacts without carefully designing the simulated metal masks. 
In the future, we will investigate how to create a good simulated dataset to further improve the network performance on real clinical CT images.

The previous prior-image-based MAR methods would utilize tissue processing to post-process the generated prior image. Whereas in our method, we directly employ the CNN output as the prior image to guide the sinogram completion network. 
In this case, we can jointly train the PriorNet and SinoNet, and the prior image generation and sinogram completion procedures can benefit from each other.
Although some mild artifacts would remain in the generated prior image, the sinogram completion network would automatically learn how to complete the sinogram from the prior sinogram, so that our final output can remove these mild artifacts.

We have trained and evaluated our method on simulated datasets, but as shown in the experiments, our framework has strong potential to be applied in \revise{ CT images with real metal artifacts.} 
\revise{Since there is no public real projection data and we need to cooperate with CT device venders to acquire such real projection data, in the current study, we use forward projection to simulate the projection data. This is a limitation of our current work and we will evaluate the effectiveness of our method on real project data in the future.}
When applying the framework into \revise{real clinical data}, one important practice issue is how to acquire the accurate metal trace and metal masks.
Although our framework is relatively robust to the metal mask segmentation, an accurate metal segmentation would further ensure the stability of the MAR results.
Deep learning has achieved promising results in various medical image segmentation problems~\cite{ronneberger2015u,litjens2017survey,yu2017volumetric}. Incorporating deep learning-based metal segmentation or advanced metal identification algorithm~\cite{meng2010sinogram} into our framework would further improve the robustness of our method. 
Recently, some works studied how to locate the shape and location of metal objects directly from the metal-corrupted sinogram~\cite{wang2010binary,meng2010sinogram}.
These binary reconstruction works can also be integrated into our framework for better metal artifact reduction. 
Moreover, it is more interesting to investigate how to simultaneously conduct metal mask identification and metal artifact reduction in a collaborative manner.

	\section{Conclusion}
\label{sec:conclusion}
We present a generalizable image and sinogram domain joint learning  framework for metal artifact reduction in CT imaging, which integrates the merits of deep  learning and conventional MAR methods.
Our framework follows the prior-image-based sinogram completion strategy and we employ two networks to conduct prior image generalization and sinogram completion.
The whole framework is trained in an end-to-end manner so that the two networks can benefit from each other in network learning.
Our framework is trained with the \revise{simulated metal artifacts data}, while the experimental results show the strong potential of our method to handle \revise{CT images with real artifacts}.
The future works include investigating how to simultaneously conduct metal mask identification and metal artifact reduction, as well as how to perform the procedure in an unsupervised manner.

	%\section{ Acknowledgments}
	%\bigskip
	%\noindent Thank you for reading these instructions carefully. We look forward to receiving your electronic files!
	
	\bibliographystyle{IEEEtran}
	\bibliography{ref}

\begin{thebibliography}{10}
\providecommand{\url}[1]{#1}
\csname url@samestyle\endcsname
\providecommand{\newblock}{\relax}
\providecommand{\bibinfo}[2]{#2}
\providecommand{\BIBentrySTDinterwordspacing}{\spaceskip=0pt\relax}
\providecommand{\BIBentryALTinterwordstretchfactor}{4}
\providecommand{\BIBentryALTinterwordspacing}{\spaceskip=\fontdimen2\font plus
\BIBentryALTinterwordstretchfactor\fontdimen3\font minus
  \fontdimen4\font\relax}
\providecommand{\BIBforeignlanguage}[2]{{%
\expandafter\ifx\csname l@#1\endcsname\relax
\typeout{** WARNING: IEEEtran.bst: No hyphenation pattern has been}%
\typeout{** loaded for the language `#1'. Using the pattern for}%
\typeout{** the default language instead.}%
\else
\language=\csname l@#1\endcsname
\fi
#2}}
\providecommand{\BIBdecl}{\relax}
\BIBdecl

\bibitem{de1998metal}
B.~De~Man, J.~Nuyts, P.~Dupont, G.~Marchal, and P.~Suetens, ``Metal streak
  artifacts in x-ray computed tomography: a simulation study,'' in \emph{IEEE
  Nuclear Science Symposium and Medical Imaging Conference}, vol.~3.\hskip 1em
  plus 0.5em minus 0.4em\relax IEEE, 1998, pp. 1860--1865.

\bibitem{kalender1987reduction}
W.~A. Kalender, R.~Hebel, and J.~Ebersberger, ``Reduction of ct artifacts
  caused by metallic implants.'' \emph{Radiology}, vol. 164, no.~2, pp.
  576--577, 1987.

\bibitem{meng2010sinogram}
B.~Meng, J.~Wang, and L.~Xing, ``Sinogram preprocessing and binary
  reconstruction for determination of the shape and location of metal objects
  in computed tomography (ct),'' \emph{Medical physics}, vol.~37, no.~11, pp.
  5867--5875, 2010.

\bibitem{park2018ct}
H.~S. Park, S.~M. Lee, H.~P. Kim, J.~K. Seo, and Y.~E. Chung, ``Ct
  sinogram-consistency learning for metal-induced beam hardening correction,''
  \emph{Medical physics}, vol.~45, no.~12, pp. 5376--5384, 2018.

\bibitem{huang2015evaluation}
J.~Y. Huang, J.~R. Kerns, J.~L. Nute, X.~Liu, P.~A. Balter, F.~C. Stingo, D.~S.
  Followill, D.~Mirkovic, R.~M. Howell, and S.~F. Kry, ``An evaluation of three
  commercially available metal artifact reduction methods for ct imaging,''
  \emph{Physics in Medicine \& Biology}, vol.~60, no.~3, p. 1047, 2015.

\bibitem{gjesteby2016metal}
L.~Gjesteby, B.~De~Man, Y.~Jin, H.~Paganetti, J.~Verburg, D.~Giantsoudi, and
  G.~Wang, ``Metal artifact reduction in ct: where are we after four decades?''
  \emph{IEEE Access}, vol.~4, pp. 5826--5849, 2016.

\bibitem{hsieh2000iterative}
J.~Hsieh, R.~C. Molthen, C.~A. Dawson, and R.~H. Johnson, ``An iterative
  approach to the beam hardening correction in cone beam ct,'' \emph{Medical
  physics}, vol.~27, no.~1, pp. 23--29, 2000.

\bibitem{kachelriess2001generalized}
M.~Kachelriess, O.~Watzke, and W.~A. Kalender, ``Generalized multi-dimensional
  adaptive filtering for conventional and spiral single-slice, multi-slice, and
  cone-beam ct,'' \emph{Medical physics}, vol.~28, no.~4, pp. 475--490, 2001.

\bibitem{meyer2010empirical}
E.~Meyer, C.~Maa{\ss}, M.~Baer, R.~Raupach, B.~Schmidt, and M.~Kachelrie{\ss},
  ``Empirical scatter correction (esc): A new ct scatter correction method and
  its application to metal artifact reduction,'' in \emph{IEEE Nuclear Science
  Symposuim \& Medical Imaging Conference}.\hskip 1em plus 0.5em minus
  0.4em\relax IEEE, pp. 2036--2041.

\bibitem{park2015metal}
H.~S. Park, D.~Hwang, and J.~K. Seo, ``Metal artifact reduction for
  polychromatic x-ray ct based on a beam-hardening corrector,'' \emph{IEEE
  Transactions Medical Imaging}, vol.~35, no.~2, pp. 480--487, 2015.

\bibitem{zhang2018convolutional}
Y.~Zhang and H.~Yu, ``Convolutional neural network based metal artifact
  reduction in x-ray computed tomography,'' \emph{IEEE Transactions Medical
  Imaging}, vol.~37, no.~6, pp. 1370--1381, 2018.

\bibitem{mehranian2013x}
A.~Mehranian, M.~R. Ay, A.~Rahmim, and H.~Zaidi, ``X-ray ct metal artifact
  reduction using wavelet domain $ l\_ $\{$0$\}$ $ sparse regularization,''
  \emph{IEEE Transactions Medical Imaging}, vol.~32, no.~9, pp. 1707--1722,
  2013.

\bibitem{muller2009spurious}
J.~M{\"u}ller and T.~M. Buzug, ``Spurious structures created by
  interpolation-based ct metal artifact reduction,'' in \emph{Medical Imaging
  2009: Physics of Medical Imaging}, vol. 7258.\hskip 1em plus 0.5em minus
  0.4em\relax International Society for Optics and Photonics, 2009, p. 72581Y.

\bibitem{prell2009novel}
D.~Prell, Y.~Kyriakou, M.~Beister, and W.~A. Kalender, ``A novel forward
  projection-based metal artifact reduction method for flat-detector computed
  tomography,'' \emph{Physics in Medicine \& Biology}, vol.~54, no.~21, p.
  6575, 2009.

\bibitem{meyer2010normalized}
E.~Meyer, R.~Raupach, M.~Lell, B.~Schmidt, and M.~Kachelrie{\ss}, ``Normalized
  metal artifact reduction (nmar) in computed tomography,'' \emph{Medical
  physics}, vol.~37, no.~10, pp. 5482--5493, 2010.

\bibitem{wang2013metal}
J.~Wang, S.~Wang, Y.~Chen, J.~Wu, J.-L. Coatrieux, and L.~Luo, ``Metal artifact
  reduction in ct using fusion based prior image,'' \emph{Medical physics},
  vol.~40, no.~8, p. 081903, 2013.

\bibitem{zhang2013hybrid}
Y.~Zhang, H.~Yan, X.~Jia, J.~Yang, S.~B. Jiang, and X.~Mou, ``A hybrid metal
  artifact reduction algorithm for x-ray ct,'' \emph{Medical physics}, vol.~40,
  no.~4, p. 041910, 2013.

\bibitem{wang1996iterative}
G.~Wang, D.~L. Snyder, J.~A. O'Sullivan, and M.~W. Vannier, ``Iterative
  deblurring for ct metal artifact reduction,'' \emph{IEEE Transactions Medical
  Imaging}, vol.~15, no.~5, pp. 657--664, 1996.

\bibitem{wang1999iterative}
G.~Wang, M.~W. Vannier, and P.-C. Cheng, ``Iterative x-ray cone-beam tomography
  for metal artifact reduction and local region reconstruction,''
  \emph{Microscopy and microanalysis}, vol.~5, no.~1, pp. 58--65, 1999.

\bibitem{lemmens2008suppression}
C.~Lemmens, D.~Faul, and J.~Nuyts, ``Suppression of metal artifacts in ct using
  a reconstruction procedure that combines map and projection completion,''
  \emph{IEEE Transactions Medical Imaging}, vol.~28, no.~2, pp. 250--260, 2008.

\bibitem{zhang2011metal}
X.~Zhang, J.~Wang, and L.~Xing, ``Metal artifact reduction in x-ray computed
  tomography (ct) by constrained optimization,'' \emph{Medical physics},
  vol.~38, no.~2, pp. 701--711, 2011.

\bibitem{wang2018image}
G.~Wang, J.~C. Ye, K.~Mueller, and J.~A. Fessler, ``Image reconstruction is a
  new frontier of machine learning,'' \emph{IEEE Transactions Medical Imaging},
  vol.~37, no.~6, pp. 1289--1296, 2018.

\bibitem{zhang2018sparse}
Z.~Zhang, X.~Liang, X.~Dong, Y.~Xie, and G.~Cao, ``A sparse-view ct
  reconstruction method based on combination of densenet and deconvolution,''
  \emph{IEEE Transactions Medical Imaging}, vol.~37, no.~6, pp. 1407--1417,
  2018.

\bibitem{litjens2017survey}
G.~Litjens, T.~Kooi, B.~E. Bejnordi, A.~A.~A. Setio, F.~Ciompi, M.~Ghafoorian,
  J.~A. Van Der~Laak, B.~Van~Ginneken, and C.~I. S{\'a}nchez, ``A survey on
  deep learning in medical image analysis,'' \emph{Medical image analysis},
  vol.~42, pp. 60--88, 2017.

\bibitem{ronneberger2015u}
O.~Ronneberger, P.~Fischer, and T.~Brox, ``U-net: Convolutional networks for
  biomedical image segmentation,'' in \emph{International Conference on Medical
  Image Computing and Computer Assisted Intervention}.\hskip 1em plus 0.5em
  minus 0.4em\relax Springer, 2015, pp. 234--241.

\bibitem{gjesteby2017deep}
L.~Gjesteby, Q.~Yang, Y.~Xi, Y.~Zhou, J.~Zhang, and G.~Wang, ``Deep learning
  methods to guide ct image reconstruction and reduce metal artifacts,'' in
  \emph{Medical Imaging 2017: Physics of Medical Imaging}, vol. 10132.\hskip
  1em plus 0.5em minus 0.4em\relax International Society for Optics and
  Photonics, 2017, p. 101322W.

\bibitem{ledig2017photo}
C.~Ledig, L.~Theis, F.~Husz{\'a}r, J.~Caballero, A.~Cunningham, A.~Acosta,
  A.~Aitken, A.~Tejani, J.~Totz, Z.~Wang \emph{et~al.}, ``Photo-realistic
  single image super-resolution using a generative adversarial network,'' in
  \emph{Proceedings of the IEEE Conference on Computer Vision and Pattern
  Recognition}, 2017, pp. 4681--4690.

\bibitem{ulyanov2018deep}
D.~Ulyanov, A.~Vedaldi, and V.~Lempitsky, ``Deep image prior,'' in
  \emph{Proceedings of the IEEE Conference on Computer Vision and Pattern
  Recognition}, 2018, pp. 9446--9454.

\bibitem{lehtinen2018noise2noise}
J.~Lehtinen, J.~Munkberg, J.~Hasselgren, S.~Laine, T.~Karras, M.~Aittala, and
  T.~Aila, ``Noise2noise: Learning image restoration without clean data,''
  \emph{ICML}, 2018.

\bibitem{park2017machine}
H.~S. Park, S.~M. Lee, H.~P. Kim, and J.~K. Seo, ``Machine-learning-based
  nonlinear decomposition of ct images for metal artifact reduction,''
  \emph{arXiv preprint arXiv:1708.00244}, 2017.

\bibitem{gjesteby2017reducing}
L.~Gjesteby, Q.~Yang, Y.~Xi, B.~Claus, Y.~Jin, B.~De~Man, and G.~Wang,
  ``Reducing metal streak artifacts in ct images via deep learning: Pilot
  results,'' in \emph{The 14th International Meeting on Fully Three-Dimensional
  Image Reconstruction in Radiology and Nuclear Medicine}, 2017, pp. 611--614.

\bibitem{gjesteby2018deep}
L.~Gjesteby, H.~Shan, Q.~Yang, Y.~Xi, B.~Claus, Y.~Jin, B.~De~Man, and G.~Wang,
  ``Deep neural network for ct metal artifact reduction with a perceptual loss
  function,'' in \emph{Proceedings of The Fifth International Conference on
  Image Formation in X-ray Computed Tomography}, 2018.

\bibitem{liang2019metal}
K.~Liang, L.~Zhang, H.~Yang, Y.~Yang, Z.~Chen, and Y.~Xing, ``Metal artifact
  reduction for practical dental computed tomography by improving
  interpolation-based reconstruction with deep learning,'' \emph{Medical
  Physics}, vol.~46, no.~12, pp. e823--e834, 2019.

\bibitem{liao2019adn}
H.~Liao, W.-A. Lin, S.~K. Zhou, and J.~Luo, ``Adn: Artifact disentanglement
  network for unsupervised metal artifact reduction,'' \emph{IEEE Transactions
  Medical Imaging}, 2019.

\bibitem{lin2019dudonet}
W.-A. Lin, H.~Liao, C.~Peng, X.~Sun, J.~Zhang, J.~Luo, R.~Chellappa, and S.~K.
  Zhou, ``Dudonet: Dual domain network for ct metal artifact reduction,'' in
  \emph{Proceedings of the IEEE Conference on Computer Vision and Pattern
  Recognition}, 2019, pp. 10\,512--10\,521.

\bibitem{lyu2020dudonet}
Y.~Lyu, W.-A. Lin, J.~Lu, and S.~K. Zhou, ``Dudonet++: Encoding mask projection
  to reduce ct metal artifacts,'' \emph{arXiv preprint arXiv:2001.00340}, 2020.

\bibitem{huang2018metal}
X.~Huang, J.~Wang, F.~Tang, T.~Zhong, and Y.~Zhang, ``Metal artifact reduction
  on cervical ct images by deep residual learning,'' \emph{Biomedical
  engineering online}, vol.~17, no.~1, p. 175, 2018.

\bibitem{wang2018conditional}
J.~Wang, Y.~Zhao, J.~H. Noble, and B.~M. Dawant, ``Conditional generative
  adversarial networks for metal artifact reduction in ct images of the ear,''
  in \emph{International Conference on Medical Image Computing and Computer
  Assisted Intervention}.\hskip 1em plus 0.5em minus 0.4em\relax Springer,
  2018, pp. 3--11.

\bibitem{isola2017image}
P.~Isola, J.-Y. Zhu, T.~Zhou, and A.~A. Efros, ``Image-to-image translation
  with conditional adversarial networks,'' in \emph{Proceedings of the IEEE
  conference on computer vision and pattern recognition}, 2017, pp. 1125--1134.

\bibitem{liao2019generative}
H.~Liao, W.-A. Lin, Z.~Huo, L.~Vogelsang, W.~J. Sehnert, S.~K. Zhou, and
  J.~Luo, ``Generative mask pyramid network for ct/cbct metal artifact
  reduction with joint projection-sinogram correction,'' in \emph{International
  Conference on Medical Image Computing and Computer Assisted
  Intervention}.\hskip 1em plus 0.5em minus 0.4em\relax Springer, 2019, pp.
  77--85.

\bibitem{yan2018deep}
K.~Yan, X.~Wang, L.~Lu, L.~Zhang, A.~P. Harrison, M.~Bagheri, and R.~M.
  Summers, ``Deep lesion graphs in the wild: relationship learning and
  organization of significant radiology image findings in a diverse large-scale
  lesion database,'' in \emph{Proceedings of the IEEE Conference on Computer
  Vision and Pattern Recognition}, 2018, pp. 9261--9270.

\bibitem{paszke2019pytorch}
A.~Paszke, S.~Gross, F.~Massa, A.~Lerer, J.~Bradbury, G.~Chanan, T.~Killeen,
  Z.~Lin, N.~Gimelshein, L.~Antiga \emph{et~al.}, ``Pytorch: An imperative
  style, high-performance deep learning library,'' in \emph{NIPS}.

\bibitem{kingma2014adam}
D.~P. Kingma and J.~Ba, ``Adam: A method for stochastic optimization,''
  \emph{arXiv preprint arXiv:1412.6980}, 2014.

\bibitem{yu2017volumetric}
L.~Yu, X.~Yang, H.~Chen, J.~Qin, and P.~A. Heng, ``Volumetric convnets with
  mixed residual connections for automated prostate segmentation from 3d mr
  images,'' in \emph{Thirty-first AAAI conference on artificial intelligence},
  2017.

\bibitem{wang2010binary}
J.~Wang and L.~Xing, ``A binary image reconstruction technique for accurate
  determination of the shape and location of metal objects in x-ray computed
  tomography,'' \emph{Journal of X-ray science and technology}, vol.~18, no.~4,
  pp. 403--414, 2010.

\end{thebibliography}

\end{document}